\documentclass[english,journal=jacsat,manuscript=article,etalmode=truncate,maxauthors=0,layout=twocolumn,number]{achemso}
\setkeys{acs}{etalmode=truncate,maxauthors=0,articletitle=true,email=false}

\usepackage{amsmath}             
\usepackage{mathtools}             
\usepackage{amssymb}             
\usepackage{wasysym}             
\usepackage{color}               
\usepackage{bm}
\usepackage{setspace}            
\usepackage{graphicx}            
\usepackage{diagbox}

\usepackage{caption}
\usepackage{subcaption}

\usepackage{siunitx}
\newcommand{\mycomment}[1]{}

\usepackage{wrapfig}             
\usepackage[dvipsnames]{xcolor}  
\usepackage{array,booktabs,multirow}      
\usepackage{comment}
\usepackage[export]{adjustbox}
\usepackage{arydshln}
\usepackage{multirow, multicol}
\usepackage[ruled, vlined, linesnumbered]{algorithm2e}

\SectionNumbersOn    

\usepackage[font=footnotesize,labelfont=bf,labelsep=period,width=0.95\textwidth]{caption}   
\usepackage[font=scriptsize,labelfont=bf]{subcaption}                     
\usepackage{hyperref}
\usepackage[capitalize]{cleveref}
\crefname{figure}{Figure}{Figures}
\Crefname{figure}{Figure}{Figures}
\crefname{table}{Tab.}{Tabs.}
\Crefname{table}{Table}{Tables}
\crefname{equation}{Eq.}{Eqs.}
\Crefname{equation}{Eq.}{Eqs.}
\crefname{section}{Sec.}{Secs.}
\Crefname{section}{Section}{Sections}

\usepackage{xr}
\definecolor{shiv_purple}{rgb}{0.6       ,  0.19607843,  0.8}
\definecolor{shiv_blue}{rgb}{0.11764706,  0.56470588,  1.}
\definecolor{shiv_green}{rgb}{0.        ,  0.57647059,  0.23529412}
\definecolor{shiv_yellow}{rgb}{0.97647059,  0.75686275,  0.1372549}
\definecolor{shiv_orange}{rgb}{1.        ,  0.54901961,  0.}
\definecolor{shiv_red}{rgb}{0.93333333,  0.20784314,  0.18039216}
\definecolor{shiv_gray}{rgb}{0.72156863,  0.71764706,  0.73333333}

\usepackage{algcompatible}
\title[INT TRANS]{Quaternion Dirac--Coulomb--Breit Integral Transformation for Relativistic Four-Component Correlated Electronic Structure Theory}

\author{Martijn Oele}
\affiliation[University of Washington]
{Department of Chemistry, University of Washington, Seattle, WA 98195, USA}

\author{Rajat Majumder}
\affiliation[University of Washington]
{Department of Chemistry, University of Washington, Seattle, WA 98195, USA}

\author{Shiv Upadhyay}
\affiliation[University of Washington]
{Department of Chemistry, University of Washington, Seattle, WA 98195, USA}

\author{Tianyuan Zhang}
\affiliation[University of Washington]
{Department of Chemistry, University of Washington, Seattle, WA 98195, USA}

\author{Ryan A Beck}
\affiliation[University of Washington]
{Department of Chemistry, University of Washington, Seattle, WA 98195, USA}

\author{Agam Shayit}
\affiliation[University of Washington]
{Department of Physics, University of Washington, Seattle, WA 98195, USA}

\author{Lucas Visscher}
\affiliation[Amsterdam]
{Department of Chemistry and Pharmaceutical Sciences, Vrije Universiteit Amsterdam, Amsterdam, 1081 HZ, The Netherlands}

\author{Xiaosong Li}
\affiliation[University of Washington]
{Department of Chemistry, University of Washington, Seattle, WA 98195, USA}
\email{xsli@uw.edu}
\begin{document}

\twocolumn[
\begin{@twocolumnfalse}
\begin{abstract}
High-accuracy correlated four-component relativistic electronic structure methods are typically formulated in terms of integrals over molecular orbital (MO). Consequently, an efficient and scalable strategy is required to deal with the complexity of transforming relativistic two-electron integrals from the atomic orbital (AO) to the MO basis. The transformation bottleneck is particularly acute for approaches that include Breit interaction integrals, whose computational and memory demands further exacerbate the transformation cost.
To overcome this challenge, we develop a quaternion-based, AO-driven direct integral transformation scheme. The method operates on scalar AO integrals and combines quaternion density–based contractions with direct Cauchy--Schwarz screening to systematically exploit integral locality. As a result, the proposed framework substantially lowers the practical computational scaling and provides an efficient, memory-conscious, and highly parallelizable pathway for the routine inclusion of relativistic Dirac--Coulomb--Breit integrals in large-scale four-component correlated calculations.
\end{abstract}
\end{@twocolumnfalse}
]

\section{Introduction} 
\label{Sec:intro}

Correlated four-component relativistic electronic structure theory employing the frequency-independent Dirac--Coulomb--Breit Hamiltonian\cite{Johnson71_41,Mark76_243,Grant14_062504,Aucar18_044113,Li21_3388,Li22_064112,Li23_171101} provides the highest level of accuracy available in quantum chemistry short of an explicit quantum electrodynamical treatment.\cite{Schwerdtfeger10_062503,Pyykko12_371,Pyykko12_45,Lindgren13_014108,Liu14_59,Liu20_180901} At the core of correlated relativistic wave function methods, including configuration interaction (CI),\cite{Fleig08_014108,Fleig10_014108, Booth20_184103,Li24_041404,Li25_11016,Li26_4358} complete active space self-consistent field (CASSCF), \cite{Faegri96_4083, Jensen08_034109, Shiozaki15_044112, Shiozaki18_014106, Neese13_104113, Schaefer18_1235,Li19_2974,Li24_2498} density matrix renormalization group (DMRG),\cite{Reiher14_041101, Li22_5011} multireference CI (MRCI), \cite{Li20_2975,Li22_141} multireference perturbation theory (MRPT),\cite{Li22_2983,Li26_arxivmrpt} and coupled-cluster (CC) \cite{Kaldor93_137,Dyall96_8769,Kaldor01_9720,Olsen07_347, Ilias09_124116,Visscher14_041107,Gomes18_174113, Cheng18_034106,Cheng19_074102,Visscher21_5509,Li24_3408,Li24_6521,Li25_084110,Li25_104112,Li26_016101} approaches, are the two-electron integrals over atom-centered basis functions.
While atomic orbital (AO) integrals are sufficient for mean-field treatments, highly accurate correlated methods are typically formulated in the molecular orbital (MO) representation. Consequently, an integral transformation from the AO to the MO basis is required, a step that frequently constitutes a dominant computational bottleneck in large-scale relativistic electronic structure calculations.\cite{Hirao04_68,Hirao06_234110,Visscher21_5509} 

In the nonrelativistic regime, the two-electron AO-to-MO Coulomb integral transformation is typically carried out through four sequential quarter transformations, yielding a formal computational scaling of $\mathcal{O}(N_b^5)$ with respect to the number of basis functions $N_b$.\cite{Yoshimine73_449,Hirata22_224102,Dutta22_204106} To mitigate this cost, extensive efforts have been devoted to accelerating the transformation step through efficient parallelization strategies, integral localization techniques, stochastic samplings, and low-rank tensor decomposition methods.\cite{Faegri96_4083,Pulay93_213,Lee95_5674,Broer00_1176,Jong17_184111,Alavi07_143001,Nagase08_431,Alavi10_041103,Li21_4258,Hirata22_224102,Li23_6255,Li23_114119}

Relativistic MO integrals are substantially more expensive to compute, primarily due to the increase in orbital dimensionality from one-component to four-component formulations and the additional requirement to transform the computationally demanding Breit interaction integrals. Although the conventional $\mathcal{O}(N_b^5)$ integral transformation algorithm remains formally applicable in the relativistic regime, it requires representing AO integrals in two-spinor or four-spinor form, resulting in a significantly increased memory footprint. Moreover, it remains unclear how the short-ranged character of the Breit operator can be effectively exploited within conventional relativistic AO-to-MO transformation schemes.\cite{Li23_9009}

In this work, we introduce a quaternion-based integral transformation framework that operates directly on AO integrals in their fundamental scalar form. To further enhance efficiency, we develop a quaternion-density-weighted Cauchy--Schwarz screening strategy that exploits the inherent locality of the Dirac--Coulomb--Breit Hamiltonian. Together, these advances substantially reduce the computational cost and memory footprint of relativistic AO-to-MO integral transformations, enabling correlated four-component calculations on significantly larger systems than previously feasible.

\section{Theory}
\label{Sec:Theory}

In this section, the following notations are used, unless otherwise specified:
\begin{itemize}
    \item $\mu, \nu, \kappa, \lambda$: large-component atomic orbitals (AOs)
    \item $p, q, r, t$: molecular orbitals (MOs)
\item The components of the $\boldsymbol{\alpha}$ vector are defined as
\begin{equation}
    \boldsymbol
    {\alpha}_{i,J} = 
    \begin{pmatrix}
    {\bf 0}_2 & \boldsymbol\sigma_J \\
    \boldsymbol\sigma_J & {\bf 0}_2
    \end{pmatrix},
    \quad J = \{x, y, z\}\notag
\end{equation}
with the $\boldsymbol\sigma$ vector consisting of Pauli matrices. 
\begin{align}
\mathbf{I}&=\begin{pmatrix} 1 & 0\\ 0&1 \end{pmatrix},
\boldsymbol\sigma_x=\begin{pmatrix} 0 & 1\\ 1&0 \end{pmatrix}, \notag\\
\boldsymbol\sigma_y&=\begin{pmatrix} 0 & -i\\ i&0 \end{pmatrix}, \boldsymbol\sigma_z=\begin{pmatrix} 1 & 0\\ 0&-1 \end{pmatrix}.\notag
\end{align}
\end{itemize}

\subsection{Dirac--Coulomb--Breit Two-Electron Interactions}
\label{Sec:TwoElectron}

In the Coulomb gauge, the instantaneous Dirac--Coulomb--Breit two-electron operator is
\begin{align}
V_{ee} &= \sum_{i=1}^N \sum_{j>i} (g^C(i,j)+g^B(i,j))\label{eq:twoeC}\\ 
    g^C(i,j) &= \frac{1}{r_{ij}} \label{eq:Coulomb}\\
    g^B (i,j) &= - \frac{\boldsymbol{\alpha}_i \cdot \boldsymbol{\alpha}_j}{r_{ij}} + \frac{(\boldsymbol{\alpha}_i \times {\bf r}_{ij}) \cdot ( \boldsymbol{\alpha}_j \times {\bf r}_{ij}) }{2r^3_{ij}} \label{eq:breit}
\end{align}
where the Breit term, \cref{eq:breit}, includes the magnetic interaction (the first term) and an additional gauge term.
Molecular four-spinor orbitals in four-component relativistic electronic structure theory can be written as
\begin{align}
    \psi_p =
    \begin{pmatrix}
        \phi_p^{L} \\
        \phi_p^{S}
    \end{pmatrix},
\end{align}
where $\phi_p^{L}$ and $\phi_p^{S}$ denote the large and small-components of the four-spinor molecular orbital, respectively, each of which is a two-spinor.
\begin{align}
    \phi_{p}^{L} &= \sum_{\tau\in\{\alpha,\beta\}}\sum_{\mu=1}^{N_b^L} c_{\mu\tau,p}^{L} \chi_{\mu\tau}^{L} , \label{eq:Lspinor}\\
    \phi_{p}^{S} &= \sum_{\tau\in\{\alpha,\beta\}}\sum_{\mu=1}^{N_b^S} c_{\mu\tau,p}^{S} \chi_{\mu\tau}^{S} \label{eq:Sspinor}
\end{align}
where $N_b$ is the number of spatial basis functions. The large-component spinor basis can be expressed as 
\begin{align}
   \chi_{\mu\alpha}^{L} =  \begin{pmatrix}
\chi_{\mu} \\
0
\end{pmatrix}   ,  \chi_{\mu\beta}^{L} =  \begin{pmatrix}
0 \\
\chi_{\mu} 
\end{pmatrix} 
\end{align}
where $\chi_{\mu}$ is a scalar spatial basis function.

The small-component basis can be obtained from the large-component basis via the restricted kinetic balance (RKB) condition, which ensures the correct nonrelativistic limit of the positive energy states.\cite{Faegri07_book,Liu10_1679,Wolf15_book,Li21_207}
\begin{align}
    \chi_{\mu\tau}^{S} = \frac{1}{2c} \boldsymbol{\sigma}\cdot\mathbf{p}~\chi_{\mu\tau}^{L}
    \label{eq:RKB}
\end{align}
Here $c$ and $\mathbf{p}$ are the speed of light and linear momentum operator, respectively. Note that under the RKB condition in four-component formulations, the numbers of large- and small-component basis functions are equal, \emph{i.e.}, $N_b=N_b^{L} = N_b^{S}$.

\subsection{AO-to-MO Integral Transformation}
\label{Sec:Int_trans}

The transformation of \emph{nonrelativistic} two-electron integrals from the AO basis to the MO basis  can be written as,
\begin{equation}
    \left(pq \vert rt\right) = \sum_{\mu \nu \kappa \lambda} c^{*}_{p \mu} c_{q \nu} \left( \mu \nu \vert \kappa \lambda \right) c^{*}_{r \kappa} c_{t \lambda}
    \label{eq:AOtoMO}
\end{equation}

The complexity of integral transformations in four-component relativistic electronic structure theory increases substantially. While nonrelativistic MO coefficients are usually chosen to be real-valued, relativistic MO coefficients are intrinsically complex-valued due to the nature of the spin-orbit operator. The four-spinor character of the molecular orbitals further enlarges the dimensionality of the AO and MO spaces. In addition, the use of the Coulomb--Breit two-electron operator introduces an expanded set of two-electron AO integrals that must be accounted for in the transformation.

\subsection{Two-Spinor Transformation}
The conventional approach for transforming relativistic two-electron integrals in the two-spinor basis employs an $\mathcal{O}(N^5)$ algorithm, in which one AO index is contracted with MO coefficients at a time.\cite{Visscher94_120, Hirao04_68}

In the two-spinor basis, the Dirac--Coulomb two-electron integral of four-spinor orbitals becomes 
\begin{align}
    (pq|rt)&=(p^Lq^L|r^Lt^L)+(p^Lq^L|r^St^S)\notag\\
    &+(p^Sq^S|r^Lt^L)+(p^Sq^S|r^St^S)\label{eq:DCInt}
\end{align}
where we used $p^L=\phi_p^L$ and $p^S=\phi_p^S$ to simplify the notation. 

For the Gaunt interaction, we have additional integrals
{\footnotesize
\begin{align}
    (p\boldsymbol\alpha q|\cdot r\boldsymbol\alpha t)&= (p^L\boldsymbol\sigma q^S|\cdot r^L \boldsymbol\sigma t^S) + (p^L \boldsymbol\sigma q^S|\cdot r^S \boldsymbol\sigma t^L)\notag\\
    &+(p^S\boldsymbol\sigma q^L|\cdot r^L \boldsymbol\sigma t^S) + (p^S \boldsymbol\sigma q^L |\cdot r^S \boldsymbol\sigma t^L)\label{eq:GauntInt}
\end{align}
}

Similarly, the gauge integrals are
{\footnotesize
\begin{align}
    (p\boldsymbol\alpha q| r\boldsymbol\alpha t)_3&= (p^L\boldsymbol\sigma q^S| r^L \boldsymbol\sigma t^S)_3 + (p^L \boldsymbol\sigma q^S| r^S \boldsymbol\sigma t^L)_3\notag\\
    &+(p^S\boldsymbol\sigma q^L| r^L \boldsymbol\sigma t^S)_3 + (p^S \boldsymbol\sigma q^L | r^S \boldsymbol\sigma t^L)_3
    \label{eq:gaugeInt}
\end{align}
}


\noindent where subscript ``3'' denotes that the integral kernel contains $\frac{\mathbf{r}_{12} \mathbf{r}^T_{12}}{r_{12}^3}$ terms in contrast to $\frac{1}{r_{12}}$ (cf. \cref{eq:breit}).

As shown in \Cref{eq:DCInt}, \Cref{eq:GauntInt}, and \Cref{eq:gaugeInt}, while the $(p^{L}q^{L}|r^{L}t^{L})$  block contains integrals identical to nonrelativistic AO integrals, building the full relativistic operator requires additional integrals. These terms originate from the small-component contributions of the four-spinor orbitals and are computationally more demanding, owing both to the increased number of required transformations and to the substantially higher cost of evaluating the corresponding AO integrals. For detailed discussions of the mathematical formulation and computational cost of relativistic two-electron AO integrals, we refer the reader to Refs.~\citenum{Li21_3388} and \citenum{Li22_064112}.

Two-electron integrals expressed in the two-spinor basis are well suited for the $\mathcal{O}\left(N_b^5\right)$ transformation algorithm, as this approach is naturally compatible with the two-spinor form of the molecular-orbital coefficients (see \Cref{eq:Lspinor,eq:Sspinor}).
While the $\mathcal{O}\left(N_b^5\right)$ transformation algorithm is formally the lowest scaling algorithm, it also suffers from a principal drawback 
 as it is not straightforward to exploit integral sparsity during the four-step transformation procedure. This limits opportunities for reduction of the actual scaling of this algorithm.

\subsection{Quaternion-Based Transformation}\label{Sec:Quaternion}

Instead of contracting the AO integrals directly with the two-spinor MO coefficients, one may instead perform the contraction with the density matrix,
\begin{equation}
D_{\nu\mu} = \sum_i c^{*}_{\mu i} c_{\nu i},
\label{eq:Den}
\end{equation}
at the expense of formally increasing the scaling to $\mathcal{O}(N_b^6)$. A key advantage of integral transformations based on density-matrix contractions is, however, that the first-half transformation can then fully exploit the effective integral screening strategies originally developed for self-consistent field (SCF) methods. In relativistic electronic structure theory, an additional advantage is that density-matrix contractions can be naturally formulated within quaternion algebra, allowing maximal exploitation of spinor symmetry and significantly reducing both computational cost and prefactor overhead.\cite{Jensen99_6211, Visscher02_759,Li21_3388,Li22_064112}

Recently, quaternion symmetry has been introduced into the Dirac--Coulomb--Breit Hamiltonian with the optimal spin and component separation.\cite{Li21_3388,Li22_064112} This approach formulates all relevant quantities, including Fock matrices, AO integrals, and density matrices, within the Pauli-matrix quaternion representation. The density matrix in the Pauli-matrix quaternion representation becomes,
\begin{align}
\label{eq:QuaternionScatter}
\mathbf{D}_s & = \mathbf{D}_{\alpha \alpha} + \mathbf{D}_{\beta \beta} \\
\mathbf{D}_z & = \mathbf{D}_{\alpha \alpha} - \mathbf{D}_{\beta \beta} \nonumber \\
\mathbf{D}_y & = i \left(  \mathbf{D}_{\alpha \beta} - \mathbf{D}_{\beta \alpha} \right) \nonumber \\
\mathbf{D}_x & = \mathbf{D}_{\alpha \beta} + \mathbf{D}_{\beta \alpha} \nonumber
\end{align}

The AO integrals expressed in the Pauli-matrix quaternion representation are highly nontrivial, particularly for the Breit interaction, due to the underlying vector algebra of the Pauli $\boldsymbol{\sigma}$ matrices and the enforcement of the restricted kinetic balance (RKB) condition (\Cref{eq:RKB}). For detailed derivations and the final working expressions, we refer the reader to Refs.~\citenum{Li21_3388,Li22_064112}.


In the following, we will use the Dirac--Coulomb integrals to illustrate the quaternion transformation algorithm developed herein.

\subsubsection{First-half Quaternion Transformation}
Casting the first-half Dirac--Coulomb two-spinor integrals in \Cref{eq:DCInt} into the Pauli-matrix quaternion representation yields the following expressions up to the order of $c^{-2}$,
{\footnotesize
\begin{align}
	&V_{\mu\nu, rt, s}^{C(2),LL}=
	\sum_{\kappa\lambda } \Bigg\{D_{\lambda\kappa,rt,s}^{LL}[ 2(\mu\nu|\kappa\lambda)]\notag\\
	&+\frac{1}{2c^2} D^{SS}_{\lambda\kappa,rt,s}
   (\mu\nu
   |\boldsymbol\nabla\kappa\cdot\boldsymbol\nabla\lambda)   \notag\\
   &+ \frac{i}{2c^2} \sum_{K=x,y,z} D^{SS}_{\lambda\kappa,rt,K}
   (\mu\nu
   |(\boldsymbol\nabla\kappa\times\boldsymbol\nabla\lambda)_K)
    \Bigg\}
\label{eq:firsthalf}
\end{align}
}

{\footnotesize
\begin{align}
	V_{\mu\nu, rt, s}^{C(2),SS} &= \frac{1}{2c^2} \sum_{\kappa\lambda} D^{LL}_{\lambda\kappa,rt,s} (\boldsymbol\nabla\mu\cdot\boldsymbol\nabla\nu
   |\kappa\lambda)	\\
	V_{\mu\nu, rt, J}^{C(2),SS} &= \frac{i}{2c^2} \sum_{\kappa\lambda} D^{LL}_{\lambda\kappa,rt,s} ((\boldsymbol\nabla\mu\times\boldsymbol\nabla\nu)_J
   |\kappa\lambda)	\\
   \notag J&=x,y,z
\end{align}
}
{ where $(\boldsymbol\nabla\mu\cdot\boldsymbol\nabla\nu
   |\kappa\lambda)$ and $((\boldsymbol\nabla\mu\times\boldsymbol\nabla\nu)_J
   |\kappa\lambda)$ are, respectively, the dot product and the $J$-th component of the cross product of the gradients associated with centers $\mu$ and $\nu$, which together form the quaternion AO tensor.}
   
The Dirac--Coulomb Hamiltonian further contains integrals of order $c^{-4}$, arising from interactions involving only small-components, commonly referred to as the SSSS integrals:
{\footnotesize
\begin{align}
	&V^{C(4),SS}_{\mu\nu,rt,s} =
	\frac{1}{8c^4}\sum_{\lambda\kappa} \Bigg\{ D_{\lambda\kappa,rt,s}^{SS} (\boldsymbol\nabla\mu\cdot\boldsymbol\nabla\nu|\boldsymbol\nabla\kappa\cdot\boldsymbol\nabla\lambda) \notag\\
       &+ i \sum_{K=x,y,z} D_{\lambda\kappa,rt,K}^{SS} (\boldsymbol\nabla\mu\cdot\boldsymbol\nabla\nu|(\boldsymbol\nabla\kappa\times\boldsymbol\nabla\lambda)_K) \Bigg\} 
\end{align}
\begin{align}
	&V^{C(4),SS}_{\mu\nu,rt,J} = \frac{1}{8c^4}\sum_{\lambda\kappa} \Bigg\{ iD_{\lambda\kappa,rt,s}^{SS}((\boldsymbol\nabla\mu\times\boldsymbol\nabla\nu)_J|\boldsymbol\nabla\kappa\cdot\boldsymbol\nabla\lambda)\notag\\
	&- \sum_{K=x,y,z}D^{SS}_{\lambda\kappa,rt,K}((\boldsymbol\nabla\mu\times\boldsymbol\nabla\nu)_J|(\boldsymbol\nabla\kappa\times\boldsymbol\nabla\lambda)_K)
	\Bigg\} 
\end{align}
}

{ Note that, for post-Hartree--Fock calculations, only Coulomb-type MO integrals need to be constructed explicitly, since the corresponding exchange-type MO integrals can be obtained by index permutation. Consequently, exchange-type integrals are not evaluated explicitly during the AO-to-MO transformation.}

As evident from these expressions, the quaternion transformation operates on lower-dimensional {  scalar} AO integral tensors of size $N_b^4$, in contrast to the two-spinor and four-spinor formulations, which involve block-sparse tensors of dimensions $(2N_b)^4$ and $(4N_b)^4$, respectively. Note that during the first-half of the transformation, both the large- and small-component contributions ($\mathbf{D}^{LL}$ and $\mathbf{D}^{SS}$) associated with the third and fourth indices are transformed.

Employing quaternion-density contraction in the transformation yields an algorithm with an intrinsic scaling of $\mathcal{O}\left(5N_b^4 N_{\mathrm{MO}}^2\right)$, where $N_{\mathrm{MO}}$ denotes the number of molecular orbitals included in the transformation. This formal scaling is higher than that of the corresponding two-spinor–based transformation, which scales as $\mathcal{O}\left(8N_b^4 N_{\mathrm{MO}}+8N_b^3N^2_{\mathrm{MO}}\right)$. We refer to the appendix for a more extensive discussion of the scaling prefactors associated with these two algorithms.

While the quaternion density-based algorithm makes it easier to reduce memory requirements by fully exploiting the block sparsity of the tensors, its main advantage becomes apparent when additionally employing integral locality and sparsity in integral screening. As we will show in the results section, this does reduce the actual scaling to below $\mathcal{O}\left(N^5\right)$.  


\subsubsection{Second-half Two-Spinor Transformation}

After the first-half transformation, the partially transformed MO integrals are expressed in the Pauli-matrix quaternion representation with a dimension of $N_b^2N_{\mathrm{MO}}^2$. In principle, the second-half of the transformation can be carried out using the same quaternion-based algorithm. However, because integral screening is already fully utilized during the first-half transformation, the second-half transformation is better implemented using a lower-scaling approach. Therefore, in this work, the second-half of the transformation is performed using an MO-based algorithm in the two-spinor representation.

Since the partially transformed MO integrals are expressed in the Pauli-matrix quaternion representation, a conversion to the two-spinor representation is required. The component-blocked two-spinor half-transformed matrix is then constructed from the corresponding Pauli components:

{\footnotesize
\begin{align}
{\bf V}^{XY} &=\frac{1}{2}
\begin{pmatrix}
{\bf V}^{XY}_s+{\bf V}^{XY}_z &   {\bf V}^{XY}_x - i {\bf V}^{XY}_y\\
{\bf V}^{XY}_x+i{\bf V}^{XY}_y  & {\bf V}^{XY}_s-{\bf V}^{XY}_z
\end{pmatrix}\\
\notag &X,Y\in\{L,S\}.
\end{align}
}

\noindent The second-half of the transformation is carried out using two-spinor MO coefficients and two-spinor matrices. 
For the Dirac--Coulomb Hamiltonian, this step can be expressed in terms of two subsequent matrix multiplications:

{\footnotesize
\begin{align}
\left(p^Xq^X \vert rt\right) &= \sum_{\mu \nu} C^{X*}_{\mu\tau_1,p} C^{X}_{\nu\tau_2,q} \left( \mu \nu \vert rt \right),\label{eq:secondhalf}\\
\notag &\tau_1,\tau_2\in\{\alpha, \beta\}
\end{align}
}

\noindent noting the diagonal nature $(p^Xq^Y | rt) = \delta_{XY} (p^Xq^X | rt)$ of the Coulomb operator.

The computational cost of the second-half of the transformation is expected to be lower than that of the first-half, as it relies on efficient matrix–vector operations and a substantially reduced memory footprint.

\subsection{Quaternion-Density-Weighted Cauchy--Schwarz Screening}
\label{Sec:Screening}

All transformation schemes discussed above can be implemented using either a direct or an in-core algorithm. In the in-core approach, all AO integrals are loaded into memory and transformed in a single sweep, typically leveraging highly optimized linear algebra libraries. While this strategy can be computationally efficient, its memory requirements scale as $\mathcal{O}(N_b^4)$ with system size, which quickly becomes prohibitive for large basis sets. In contrast, the direct integral transformation evaluates AO integrals on the fly, significantly reducing memory demands and enabling calculations involving very large integral sets and basis expansions.

A key advantage of direct integral transformation based on quaternion integral--density contraction is its ability to exploit locality and sparsity. This capability is particularly important for relativistic integrals, which are strongly localized and dominated by nearest-neighbor interactions.\cite{Visscher02_304, Li23_9009}

A widely used screening algorithm is based on the Cauchy--Schwarz inequality, along with extensions that incorporate density matrix elements into the evaluation of integral upper bounds.\cite{Whitten73_4496,Koller83_7583, Pople94_65, Grumbach95_1456,Head-Gordon97_9708,Frisch06_104103, Ochsenfeld17_144101,Gill23_842,Barca24_10424}
For relativistic integrals, the density-weighted Cauchy--Schwarz inequality must be applied separately to each quaternion scalar integral. In the case of the Dirac--Coulomb scalar-relativistic term, the corresponding Cauchy--Schwarz bound is given by

{\footnotesize
\begin{align}
&|\left( \mu \nu \vert \boldsymbol\nabla\kappa \cdot\boldsymbol\nabla\lambda \right){D}^{SS}_{\lambda\kappa}| \notag\\
&\leq \sqrt{\left( \mu \nu \vert \mu \nu \right)} \cdot \sqrt{\left( \boldsymbol\nabla\kappa\cdot \boldsymbol\nabla\lambda \vert \boldsymbol\nabla\kappa\cdot \boldsymbol\nabla\lambda \right)} |\tilde{D}^{SS}_{\lambda\kappa}|_\text{max}
\label{eq:SchwarzDen}
\end{align}
}

\noindent where $|\tilde{D}^{SS}_{\lambda\kappa}|_\text{max}$ is the largest absolute value for all $\{r,t\}$. Similar density-weighted Cauchy--Schwarz bounds can be straightforwardly generalized to other classes of relativistic integrals.

\section{Results}
\label{Sec:Results}
Benchmark calculations were performed using a development version of the Chronus Quantum suite.\cite{Li20_e1436} The speed of light for this study was set to 137.035999074 a.u. All calculations were performed within the Kramers unrestricted four-component framework using a finite-nucleus Gaussian distribution.\cite{Dyall97_207,Li21_207} All post Hartree-Fock calculations were performed within the no-pair approximation, where the molecular orbital space is restricted to the positive energy states.\cite{Faegri07_book} The MO-driven quaternion-integral transformation that was used to compare with the current implementation was first published in Ref. \citenum{Li23_044101}. To compare the quaternion-based AO direct algorithm with a two-spinor in-core $\mathcal{O}\left(N_b^5\right)$ integral transformation, a modified version of PySCF was used.\cite{Chan20_024109,Sun25_016101} The integral transformations used the LIBCINT library for integral evaluation.\cite{Sun15_1664} The uncontracted Dyall basis sets were used for all calculations.\cite{Dyall04_403, Dyall07_491} These were performed on the Hyak HPC managed by the University of Washington and Bridges-2 managed by the Pittsburgh Supercomputing Center.

\subsection{Effectiveness of the Density-Weighted Cauchy--Schwarz Screening}
\label{Sec:hg_screening}
\begin{figure}[ht]
    \centering
    \includegraphics[width=\linewidth]{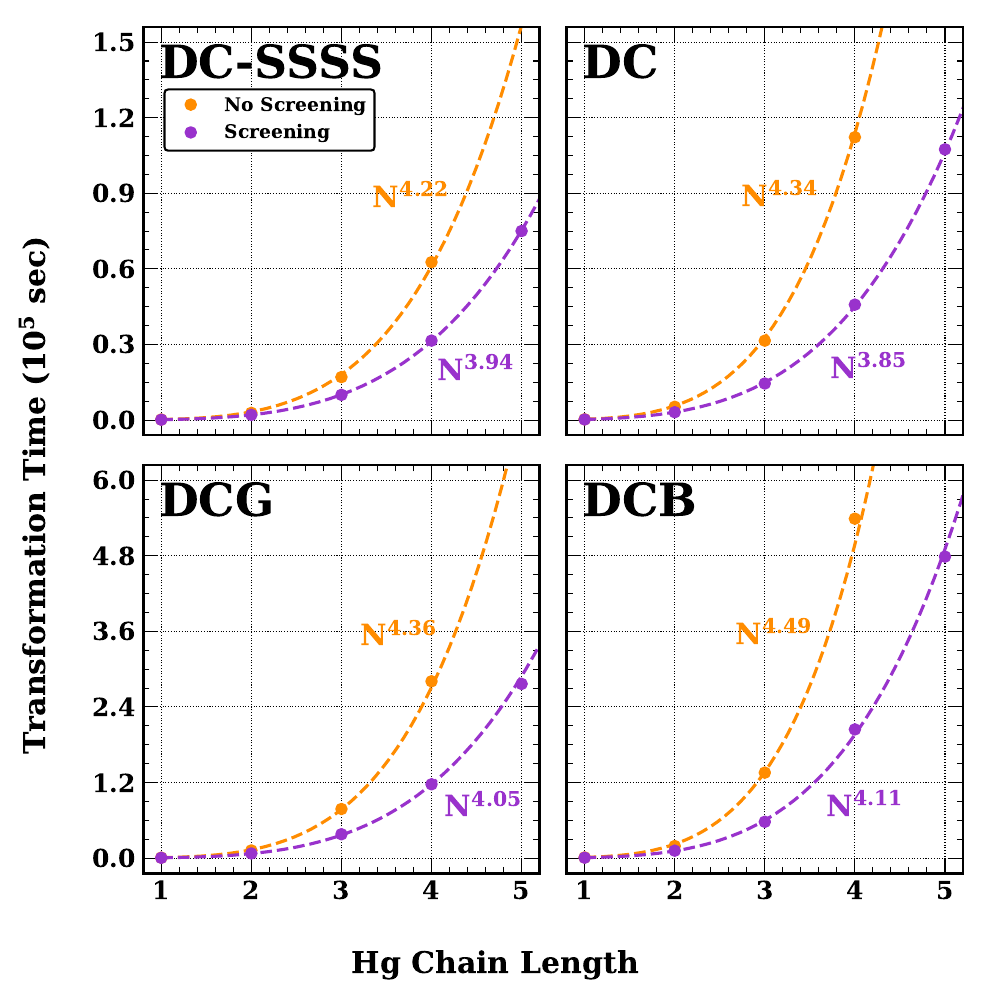}
    \captionsetup{width=\linewidth}
    \caption{Integral transformation time (sec) as a function of the number of mercury (Hg) atoms in a linear chain with and without the use of quaternion-density-weighted Cauchy--Schwarz screening. The graph titles refer to the Dirac--Coulomb Hamiltonian without the SSSS-type integrals (DC-SSSS), the full Dirac--Coulomb Hamiltonian (DC), the Dirac--Coulomb--Gaunt Hamiltonian (DCG), and the Dirac--Coulomb--Breit Hamiltonian (DCB). A Schwarz threshold of $10^{-12}$ was used for the screened calculations. A distance of 3.05 \AA\  was used between the mercury atoms. A growing active space with 18 orbitals per mercury atom was used. The Dyall-ae2z basis set was used with $N_b=213$ per mercury atom. The dashed lines represent the monomial (power-law) fit to the data.}
    \label{Fig:Hgchain_screening}
\end{figure}

A key advantage of the quaternion-based AO direct integral transformation is its ability to efficiently exploit density-weighted Cauchy--Schwarz screening. This capability is illustrated in \Cref{Fig:Hgchain_screening}, which reports integral transformation timings for a series of linear mercury (Hg) chains with an internuclear separation of 3.05~\AA~determined from experimental X-ray diffraction.\cite{Wagner66_1259}

All calculations employed the Dyall-ae2z basis set, for which each mercury atom contributes $N_b=213$ spatial AO basis functions. The MO transformation space was systematically expanded with system size, with each mercury atom contributing 18 orbitals (5d, 6s, and 6p) to the active space. A Schwarz threshold of $10^{-12}$ was used in the quaternion-based AO direct integral transformations. Calculations were performed on the University of Washington's Hyak high-performance computing system using 192 AMD EPYC 9354P cores and up to 1,536~GB of memory.

As shown in \Cref{Fig:Hgchain_screening}, the use of quaternion-density--weighted Cauchy--Schwarz screening leads to a substantial reduction in computational wall time, with the benefit becoming increasingly pronounced as the length of the mercury chain grows. For Hg$_2$, approximately 70\% of AO shells are screened out, and this fraction increases to 84\% for Hg$_3$, illustrating the enhanced effectiveness of screening for larger systems.

\Cref{Fig:Hgchain_screening} also shows the monomial scalings of the quaternion-based AO direct integral transformation algorithms, obtained via a least-squares fit. Formally, the conventional density-based integral transformation scales as $\mathcal{O}\left(N_b^6\right)$. The quaternion formalism, based on scalar integrals, substantially reduces the effective scaling by exploiting quaternion symmetry and low-prefactor contractions in the spatial basis. As a result, even in the absence of integral screening, the algorithm exhibits an effective scaling of $\mathcal{O}(N_b^{4.22\text{–}4.49})$ for the four types of relativistic Hamiltonians considered here. 

Including quaternion-density-weighted Cauchy--Schwarz screening further reduces the effective scaling to $\mathcal{O}(N_b^{3.85\text{–}4.11})$.
We also note that the largest reduction due to integral screening is observed for DC with the SSSS contribution. This behavior is expected, as the $\left( \mu^S \nu^S \vert \kappa^S \lambda^S \right)$ integrals have intrinsically small magnitudes due to their $c^{-4}$ prefactor.


\subsection{Computational Complexities of Different AO-to-MO Integral Transformation Algorithms}
\begin{figure}[h]
    \centering
    \includegraphics[width=\linewidth]{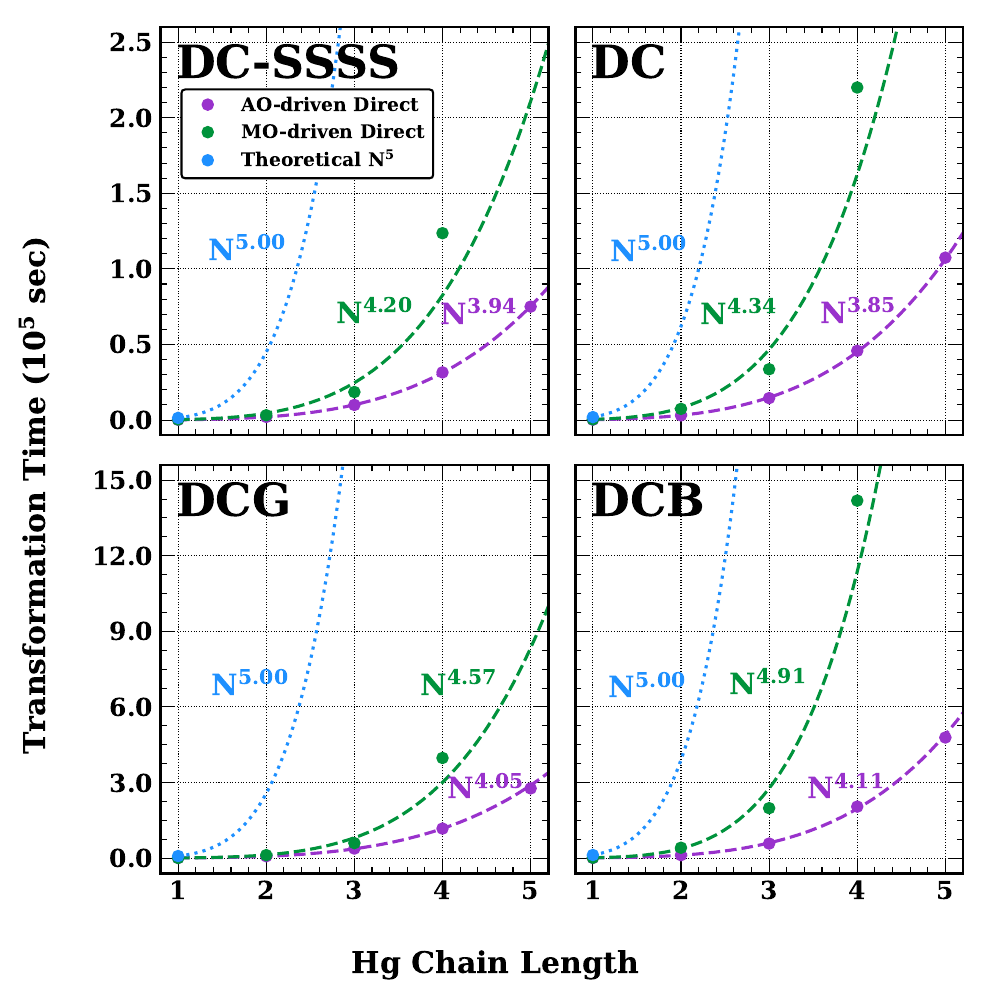}
    \captionsetup{width=\linewidth}
    \caption{Integral transformation time (sec) as a function of the number of mercury (Hg) atoms in a linear chain using different two-electron integral transformation algorithms. The graph titles refer to the Dirac--Coulomb Hamiltonian without the SSSS-type integrals (DC-SSSS), the full Dirac--Coulomb Hamiltonian (DC), the Dirac--Coulomb--Gaunt Hamiltonian (DCG), and the Dirac--Coulomb--Breit Hamiltonian (DCB). The direct transformations used a Schwarz threshold of $10^{-12}$. {   The theoretical $N^5$ transformation is shown to guide the eye, with the prefactor computed from a single point.} A distance of 3.05 \AA\ was used between the mercury atoms. An active space of 18 orbitals per mercury atom was used for all calculations. The Dyall-ae2z basis set was used with $N_b=213$ per mercury atom. The dashed and dotted lines represent the monomial (power-law) fit to the data.}
    \label{Fig:hgchain}
\end{figure}

Direct contraction schemes can be classified as either AO-driven or MO-driven. In an MO-driven integral transformation, the MO indices define the outermost loop, so the AO integrals associated with a given MO block are constructed on demand; when the outer MO index changes, these AO integrals must be recomputed. By contrast, an AO-driven integral transformation places the AO indices in the outermost loop, allowing the corresponding MO integrals to be partially transformed and accumulated progressively without redundant recomputation. {  The AO-driven integral transformation introduced in this work exploits this structure for parallelization, distributing the first two AO indices across MPI ranks and the last two across the number of threads (cores).}

\Cref{Fig:hgchain} compares the computational cost of the three integral transformation algorithms considered in this work: a quaternion-based AO-driven direct algorithm, a quaternion-based MO-driven direct algorithm, and a two-spinor-based in-core algorithm. No spatial symmetries were exploited in either the AO or MO integrals.
Both direct approaches employ the quaternion-based integral transformation with a formal scaling of
$\mathcal{O}\left(N_b^4 N_{\mathrm{MO}}^2\right),
$ whereas the in-core two-spinor transformation has a theoretical time complexity of $\mathcal{O}\left(N_b^4 N_{\mathrm{MO}}\right)$.

For the in-core transformation, timing data could only be obtained for a single mercury atom, owing to the prohibitive memory requirements associated with storing all two-spinor AO integrals. Based on this single data point, an estimated trend line was constructed by fixing the scaling to the theoretical $\mathcal{O}\left((2N_b)^4 N_{\mathrm{MO}}\right)$ behavior. Note that this estimate is only intended to guide the eye, as the prefactor cannot be acruatly determined from a single data point. Readers are therefore not advised to interpret the estimate quantitatively or extrapolate it beyond the qualitative scaling trend illustrated here.

From \Cref{Fig:hgchain}, the quaternion-based direct transformation exhibits significantly lower computational cost and a reduced observed scaling compared to the two-spinor-based in-core integral transformation. This improvement is primarily attributed to the use of scalar integrals and an efficient screening strategy.

Comparing the AO-driven and MO-driven quaternion-based transformations, the AO-driven approach clearly exhibits a lower computational scaling. This advantage arises because the MO-driven transformation incurs substantial redundancy by recomputing AO integrals for each block of MO integrals. In contrast, the AO-driven formulation eliminates these redundant computations entirely, leading to a significant performance improvement.
Notably, the largest reduction in computational cost is observed for the DCB Hamiltonian, substantially extending the practical applicability of these calculations to larger system sizes.

\subsection{Large-Scale Two-Electron Transformations}
\label{Sec:Mass}
\subsubsection{Linear Plutonium (Pu) Chain}
\begin{figure}[h]
    \centering
    \includegraphics[width=\linewidth]{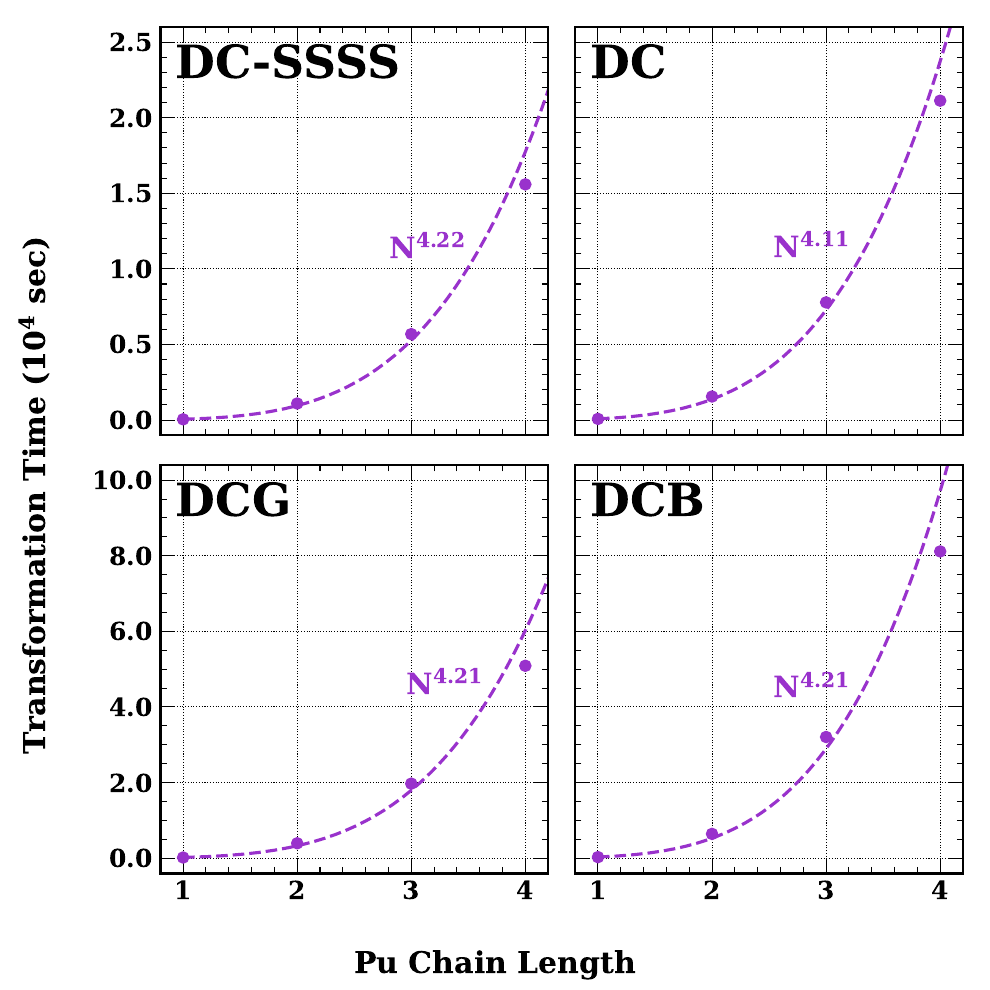}
    \captionsetup{width=\linewidth}
    \caption{Integral transformation time (sec) as a function of the number of plutonium (Pu) atoms in a linear chain. The graph titles refer to the Dirac--Coulomb Hamiltonian without the SSSS-type integrals (DC-SSSS), the full Dirac--Coulomb Hamiltonian (DC), the Dirac--Coulomb--Gaunt Hamiltonian (DCG), and the Dirac--Coulomb--Breit Hamiltonian (DCB). A distance of 4.375 \AA\ was used between the Pu atoms. An active space with 16 orbitals per plutonium atom was used for the calculations. The Dyall-v2z basis set was used, comprising $N_b=282$ spatial AO basis functions for each plutonium atom. The dashed lines represent the monomial (power-law) fit to the data.}
    \label{Fig:Pu_chain}
\end{figure}

To demonstrate the efficiency and hybrid parallelism of the newly developed integral transformation, a linear plutonium (Pu) chain was selected as the model system. An interatomic distance of 4.375~\AA\ was adopted,\cite{Ray99_5105} together with a Schwarz screening threshold of $10^{-12}$. All calculations employed the Dyall-v2z basis set,\cite{Dyall02_335} in which each plutonium atom contributes $N_b=282$ spatial AO basis functions. 

The monomial fits obtained by least-squares minimization of the scaling error yield effective scaling exponents of 4.22, 4.11, 4.21, and 4.21 for the DC-SSSS, DC, DCG, and DCB Hamiltonians, respectively. These observed scalings are substantially lower than the formal $\mathcal{O}(N_b^6)$ behavior and can be attributed to the use of quaternion-based integral transformation schemes combined with an efficient Schwarz screening strategy. The impact of screening is further illustrated in \Cref{Tab:screening_pu}, where approximately 90\% of AO shells are screened out for the Pu$_4$ chain. Notably, SSSS-type integrals exhibit an even higher screening rate, consistent with trends observed in our prior studies.

\begin{table}
\footnotesize
    \centering
    \begin{tabular}{ccccl}
    \hline
         Chain length&  DC-SSSS &   SSSS & Gaunt & gauge\\
         \hline
         1&  14.1 \%&  35.7 \%&  18.8 \%&19.4 \%\\
         2&  67.6 \%&  80.3 \%&  69.3 \%&69.4 \%\\
         3&  82.8 \%&  90.6 \%&  83.6 \%&83.5 \%\\
         4&  89.4 \%&  94.6 \%&  89.9 \%&89.8 \%\\
         \hline
    \end{tabular}
    \captionsetup{width=\linewidth}
    \caption{Percentage of the AO shells screened out by the density-weighted Schwarz screening scheme for the linear plutonium chain. The length of the atomic chain and the two-electron integral classes involved are mentioned in the column header. }
    \label{Tab:screening_pu}
\end{table}

To better understand the computational time complexity of the new integral transformation, we analyzed the wall times associated with individual components of the algorithm. The implementation can be decomposed into four stages: (i) computation and construction of quaternion scalar integrals, (ii) quaternion density formation, (iii) the first-half quaternion integral--density contraction, and (iv) the second-half transformation involving contraction with MO coefficients.

As shown in \Cref{Tab:sep_pu}, the first-half quaternion integral--density contraction is the most time-intensive stage. This behavior is expected, as this step formally exhibits $\mathcal{O}(N_b^6)$ scaling and therefore dominates the overall computational cost. The second most expensive stage is the second-half contraction with the MO coefficients, which formally scales as $\mathcal{O}(N_b^5)$. In contrast, the evaluation and construction of quaternion AO integrals contribute only a minor fraction of the total computational cost, except for the gauge integral terms, which constitute a comparatively expensive component of the overall workflow.

\begin{table*}
    \centering
    \begin{tabular}{ccccl}
    \hline
         &  DC-SSSS&  SSSS& Gaunt &gauge\\
         \hline
         Quaternion AO integrals&  \num{9.13}&  \num{21.89}&  \num{18.68}&\num{253.4}\\
         Quaternion Density &  \num{61.49}&  \num{31.1}&  \num{61.88}&\num{60.67}\\
         1/2 transform&  \num{2969.0}&  \num{804.8}&  \num{9215.0}&\num{9556.0}\\
         2/2 transform &  \num{24.85}&  \num{17.48}&  \num{35.22}&\num{34.91}\\
         \hline
    \end{tabular}
    \caption{Compute time (in seconds) for the plutonium triatomic chain using the newly implemented quaternion AO-driven direct integral transformation. The maximum wall time across all threads is reported. ``Quaternion AO integrals'' denotes the time required for integral evaluation with the LIBCINT library and the construction of quaternion AO integrals. ``Density'' corresponds to the time for density matrix construction. ``1/2'' represent the first-half of the quaternion transformation, while ``2/2'' denotes the second-half transformation. The two-electron integral classes involved are mentioned in the column header.}

    \label{Tab:sep_pu}
\end{table*}

\subsubsection{Planar Au\textsubscript{6} Cluster}
Finally, we demonstrate the capabilities of the quaternion-based AO direct integral transformation algorithm by performing a DCB calculation on a planar Au$_6$ cluster. Gold clusters have attracted sustained interest due to their relevance in quantum information science, biomedicine, and nanocatalysis.\cite{Haruta20_464,Vermonden22_04,Yang23_468,Sun22_105022,Li23_4666} Accurate characterization of their electronic structure is particularly challenging, as relativistic effects play an important role in governing their bonding and spectroscopic properties.\cite{Pal09_7101,Li23_4666} The Au$_6$ geometry was obtained by optimizing the structure using the def2-TZVP basis set\cite{Alrichs05_3297} in conjunction with the PBE0 hybrid density functional\cite{Burke96_9982,Barone99_6158} augmented with D3 dispersion corrections,\cite{Grimme11_211,Bannwarth16_5105} as implemented in the Gaussian~16 software package.\cite{G16A03}

Subsequent relativistic four-component calculations employed the Dyall-ae2z basis set, which yields $N_b=1278$ spatial AO basis functions for the Au$_6$ cluster. Such a problem size is prohibitive for an in-core integral transformation: storing the full two-spinor four-index AO integral tensor would require approximately 39 TB of memory. The quaternion AO-driven integral transformation circumvents this bottleneck and enables large-scale parallel execution.  An MO transformation space of 48 spinors was utilized. The Au$_6$ calculation reported here was performed on the Bridges-2 system at the Pittsburgh Supercomputing Center, using 2048 CPU cores (16 nodes, 128 cores per node) on AMD EPYC 7742 processors, with 256~GB of memory available per MPI node.

The quaternion-based AO direct transformation of the four-component DCB two-electron integrals required only 14.0 hours of wall time, enabling subsequent high-level correlated calculations, such as relativistic four-component coupled-cluster and configuration interaction methods, to be performed for molecular systems of this size.

\begin{figure}[h]
    \centering
    \includegraphics[width=0.5\linewidth]{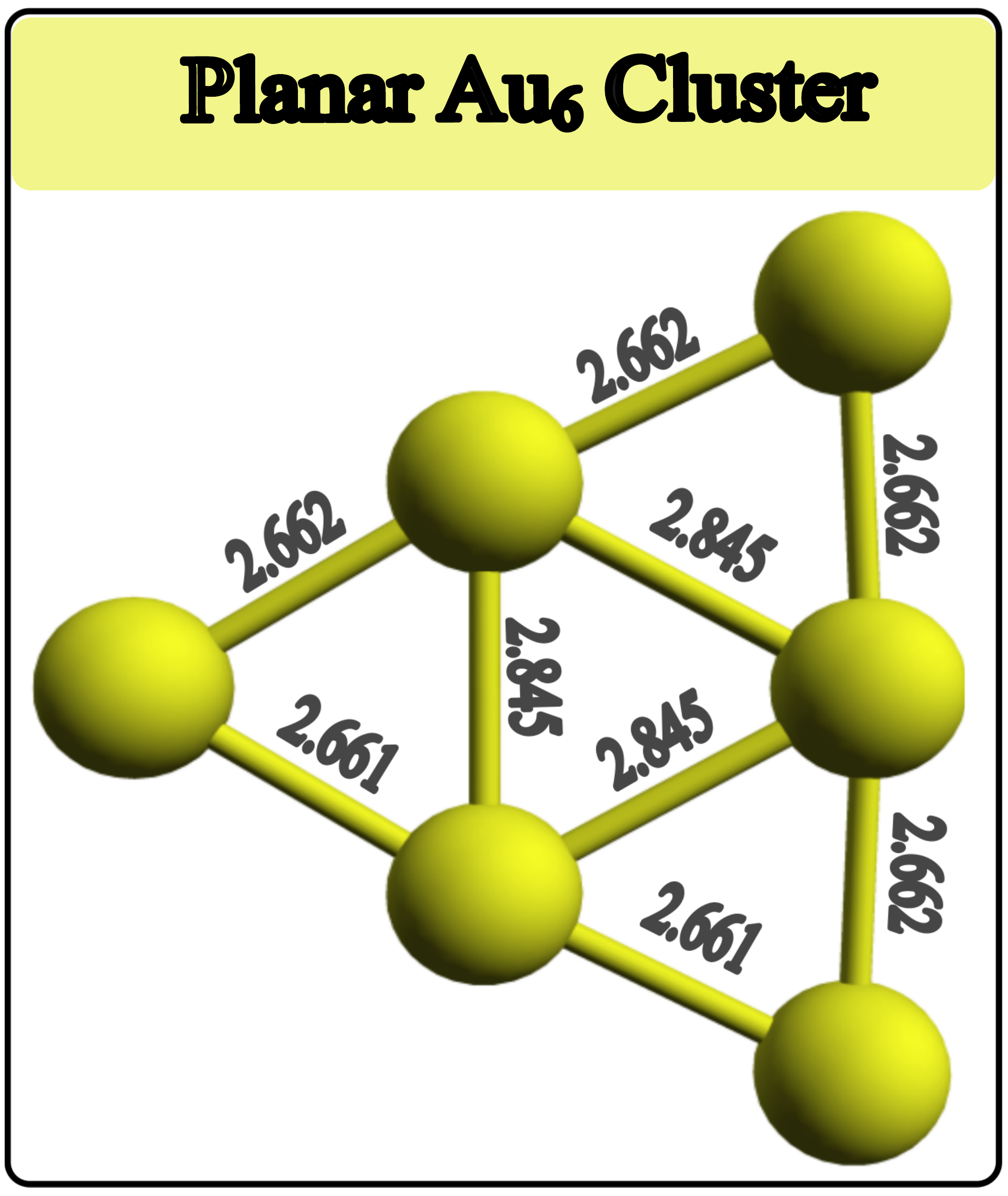}
    \captionsetup{width=\linewidth}
    \caption{Structure of the planar Au\textsubscript{6} cluster used for demonstrating large parallelization capabilities of the quaternion four-component two-electron integral transformation algorithm. The bond distances ($\text{\AA}$) between individual Au atoms are labeled in the figure.}
    \label{Fig:Au6}
\end{figure}

\section{Conclusion}
\label{Sec:Conclusion}
To enable correlated four-component relativistic treatments of large molecular systems, an efficient, scalable, and robust AO-to-MO Dirac--Coulomb--Breit integral transformation is essential. In this work, we introduce a quaternion-based relativistic integral transformation framework that operates directly on atomic orbital integrals in their fundamental scalar form. The first-half transformation employs density contraction with component and spin separation in the Pauli-matrix quaternion representation, enabling quaternion density-weighted Cauchy--Schwarz screening to be applied naturally and efficiently to exploit integral locality at an early stage. The second-half transformation utilizes a two-spinor contraction with molecular orbital coefficients, further reducing the overall computational cost.

Numerical benchmarks demonstrate that the proposed algorithm efficiently performs Dirac--Coulomb--Breit integral transformations for large four-component relativistic systems that were previously computationally inaccessible. By combining a quaternion-based formulation with an efficient quaternion density-weighted Cauchy--Schwarz screening strategy, the method achieves a reduced practical scaling compared to conventional transformation approaches.

\begin{acknowledgement}
The development of variational relativistic post-SCF methods is supported by the U.S. Department of Energy, Office of Science, Basic Energy Sciences, in the Computational and Theoretical Chemistry program (Grant No. DE-SC0006863). The development of the Chronus Quantum computational software is supported by the Office of Advanced Cyberinfrastructure, National Science Foundation (Grant No. OAC-2103717). The authors acknowledge the use of facilities and instrumentation supported by the U.S. National Science Foundation through the UW Molecular Engineering Materials Center (MEM-C), a Materials Research Science and Engineering Center (DMR-2308979).
\end{acknowledgement}

\section*{Data Availability}
The data that support the findings of this study are available within the article.

\section*{Appendix}
    The scaling prefactors for the $\mathcal{O}(N^5)$ transformation of an RKB basis are computed according to the scheme implemented in Ref. \citenum{Visscher21_5509}. Transforming the Coulomb $LLLL$ and $LLSS$ contributions to an $LLMM$  intermediate representation (with $M$ indicating the molecular basis) requires $(4\times2\times2)N_b^4N_{MO} + (4\times2\times2)N_b^3N^2_{MO}$ floating point operations. Here the individual prefactors concern respectively: use of complex algebra, separate $\alpha\alpha$ and $\beta\beta$ spin integration, separate $LL$ and $SS$ integration. The $LLLL$ and $LLSS$ contributions to the $LLMM$ integrals can be added prior to second-half transformation, which then requires an additional number of $(4\times2)N_b^2N^3_\text{MO} + (4\times2)N_bN^4_\text{MO}$ floating point operations. The transformation of the $LLSS$ and $SSSS$ integral contributions has the same scaling. The transformation of Gaunt and gauge integrals is not implemented in Ref. \citenum{Visscher21_5509}, but following a similar scheme would yield higher prefactors due to appearance of the Pauli matrices in the kernels. 
    
    The scaling for the first-half transformation of the Coulomb integral contributions can be contrasted with the current approach in which a density matrix is contracted with the AO integrals. This one-step procedure, \cref{eq:firsthalf}, to obtain $LLMM$ has a scaling of $5\times N_b^4N_{MO}^2$ with the prefactor arising from the quaternion decomposition of the integral and density matrix into scalar and vector terms. If we transform all MOs, \emph{i.e.}, $N_\text{MO}=2N_b$ we have a prefactor of 96 for the $\mathcal{O}\left(N_b^5\right)$ algorithm and 20 for the $\mathcal{O}\left(N_b^6\right)$ algorithm. This is, however, not very realistic as typically only a subset of MOs is used in correlated calculations. Setting $N_\text{MO}=\frac{1}{2}N_b$ instead yields a prefactor of 12 for the $\mathcal{O}\left(N_b^5\right)$ algorithm and 1.25 for the $\mathcal{O}\left(N_b^6\right)$ algorithm. 

\newpage 

\bibliography{
    references/Journal_Short_Name,
    references/Li_Group_References, 
    references/CI,
    references/sparseDAS,
    references/mcscf,
    references/rel_other,
    references/CI_approx,
    references/other_correlated_methods,
    references/DCB,
    references/RelCC
} 

\end{document}